\documentclass[preprint,10pt,flushrt]{aastex}

\usepackage{natbib}
\usepackage{graphics}
\usepackage{amssymb,amsmath}
\usepackage{graphicx}
\usepackage{mathrsfs}
\newcommand{\ld}{\lambda/D}

\begin{abstract}
The past five years has seen a surge in research and innovative ideas for the imaging of extrasolar planets, particular terrestrial ones.  We expect that within the next decade a space observatory will be launched with the objective of imaging earthlike planets.  Because of the limited lifetime of such a mission and the large number of potential targets, integration time is a critical parameter.  In fact, integration time is the primary metric in evaluating various design approaches for the high contrast imaging system.   In this paper we present a new approach to determining the existence of a planet in an observed system using Bayesian hypothesis testing.  Rather than perform photometry, or rely on vision to determine the existence of a planet, this approach evaluates the image plane data statistically under certain assumptions about the prior probability distributions.  We show that extremely high confidence can be achieved in substantially shorter integration times than conventional photometric methods.
\end{abstract}

\begin{document}

\title{Bayesian Hypothesis Testing for Planet Detection}
\author{Isabelle Braems\footnote{I. Braems received a fellowship from the Institut National de Recherche en Informatique et Automatique (INRIA) }} 
\affil{Laboratoire d'Etudes des MatŽriaux Hors-Equilibre (LEMHE)} 
\author{N. Jeremy Kasdin}
\affil{Princeton University}
\date{}
\maketitle

\section{Introduction}

The discovery of more than 100 extrasolar Jupiter-sized planets in just the last decade has generated enormous interest, both among astronomers and the public, in the problem of discovering and characterizing Earthlike planets.  NASA is already planning its next large space-based observatory, the \emph{Terrestrial Planet Finder (TPF)}, with a planned launch date toward the end of the next decade.  TPF's primary objective will be to discover Earthlike planets and characterize them for indications of life.  

The technical challenges for TPF are great.  Foremost among them is the problem of high-contrast imaging.  In order to discover as many planets as possible, it is necessary to design an imaging systems that achieves very high contrast between the parent star and the planet as close as possible to the star.  An earlier study by \citet{ref:Brown} 
indicates that a $D = 4$m class visible-light 
(i.e. $400\text{nm} \le \lambda \le 650\text{nm}$)
instrument ought to be able to discover
about 50 extrasolar Earth-like planets if it can provide contrast
of $10^{-10}$ at an angular separation of $3\ld$ and that a $4 \times 10$m class
telescope ought to be able to discover about 150 such
planets if it can provide the same contrast at a separation of $4\ld$.

There are myriad of methods being proposed to achieve the high contrast necessary for planet detection.  Our  interest has been in shaped pupil coronagraphy.  Many of  our optimal approaches are described in \cite{ref:KVSL, VSK02, VSK03}.  Of particular importance is the discussion in \cite{ref:KVSL} on performance metrics used to compare and optimize various coronagraphic approaches.  The key factor here is integration time.  All planet finding missions must minimize the integration time in order to provide the most possible observations and the least sensitivity to instabilities in the telescope and spacecraft.  To that end, quantitative measures of integration time, derived for specific image processing techniques, are essential.

In \cite{ref:KVSL}
 we discussed a number of different approaches
for planet detection, each with a corresponding integration time
to reach a specified signal-to-noise ratio ($S/N$).  The objective
was to find quantitative metrics that could be used to compare
different coronagraph approaches.  The simplest detection
integration time threshold,  $t_1$, was proposed such that for
$\tau \geq t_{1}$ the decision of whether a planet exists can be
taken by a human eye. This threshold is based on the expected
signal to noise ratio $S/N$
\begin{equation}
S/N = \frac{I_p\Delta P }{\sqrt{I_b \Delta
S}}\sqrt{t_1}\label{eqsn},
\end{equation}
where it is assumed that the planet intensity $I_p$ is known, the
background is uniform and its intensity $I_b$ is also know and
$\Delta P$ is the area under the PSF $P$ in the integration area
$\Delta S$ around the planet location to be denoted $\zeta^*$, where $\zeta^*$ refers to the (continuous) location in the image plane of the central peak of the planet's PSF
(note that $\zeta^*$ can be considered a member of $\mathbb R^2$,
thus representing the two-dimensional position of the planet on
the image plane).  The usual assumption is that the human eye can
detect a planet when the $S/N$ ratio exceeds 5, setting the typical detection criterion.  Note the dependence on PSF shape; as the coronagraphs we are considering operate by modifying the shape of the PSF, there is some sensitivity of the integration time to the specific coronagraph design.  \cite{ref:Brown_Burrows} call this ratio of PSF area to pixel area the ``sharpness''.

Unfortunately, this photometric approach to detection is limited, as it
relies on an empirical criterion that does not provide any
information about the confidence level to be associated with the
collected images. In \cite{ref:KVSL},
other integration times were computed but they were based on the
assumption that photometry was performed on the image in an
effort to estimate the planet (and perhaps background) irradiance.
The goal of this paper is to
establish a simple detection criterion to be used before any
photometry analysis in order to allow a smaller integration time,
while providing quantitative information on the confidence
associated with a planet decision.  We also plan to study modifications of the coronagraph design optimization using these results.

The basic principal is to use probabilistic hypothesis testing to
confirm or deny the hypothesis that a planet is located at a
given pixel.  Among several hypothesis testing techniques,
Bayesian approaches seem promising. Their appeal stems, in large
part, from the fact that these tests are conditioned on the
collected data, no matter the amount, while \emph{frequentist}
approaches require interpolation on fictitious data (as multiple
data sets for determining statistics will not be available).
Moreover, the mathematical framework of Bayesian methods includes
marginalization, \emph{i.e.}, an easy way to take into account
uncertain nuisance parameters (see \cite{Berger}). For these
reasons and others, Bayesian techniques have resulted in a significant
breakthrough in the last decade (see for example \cite{Loredo},
\cite{Loredo2}, \cite{Defay} and \cite{Aigrain}).

For the sake of simplicity and demonstration, we only consider 1D images in this work (that is, the dependence of the PSF along only one axis of the image plane).
While the
real image is a continuous function of the position $\zeta$: $%
\mathcal{F}\left( \zeta\right) =I_{b}+I_{p}P\left( \zeta-\zeta^*
\right)$, the CCD collects a photon noise corrupted discretized
image $Z=\left(z_{n}\right)_{n=1,\ldots,M}$ where $z_{n}$ stands
for the intensity of the $n-th$ pixel. We use $n^*$ to indicate the index of
the pixel containing $\zeta^*$: $\zeta^* \in [n^*\Delta
\alpha,(n^*+1)\Delta \alpha]$. Due to the low photon count at
these short integration times, each $z_{n}$ is modeled as a
Poisson random variable whose mean is
\begin{equation}
\lambda_{n}(I_b,I_p,\zeta^*)=\tau \left(I_{b}\Delta
\alpha+I_{p}\int_{\zeta_n}^{\zeta_{n+1}}P\left( \zeta -\zeta^*
\right) d\zeta\right)\triangleq \tau\left(I_{b}\Delta \alpha
+I_{p}\Delta P_{n}\left( \zeta^* \right) \right), \forall
n=1,...,M. \label{eq:PoissonRate}
\end{equation}
where $\Delta \alpha$ is the size of a
pixel and $(\zeta_n,\zeta_{n+1})$ refer to the location of the edges of pixel $n$.  In this preliminary study, we will assume that
$I_{b}$ is uniform, we will neglect the speckle noise due to imperfect optics, and we
will consider only monochromatic signals.  Clearly these are
rather restrictive assumptions, but they allow us to focus
attention on the technique and compare our results to the
previous one.  We hope during the next few years to expand this
effort and use the same Bayesian ideas on the more realistic polychromatic
system with speckle.  For instance, a planet might be separated
from speckle by incorporating into the model a few discrete
frequencies (most likely via a dichroic) or via multiple
measurements at different observatory orientations.

\section{Detection criterion}
 In most cases
(except \cite{Loredo2}) Bayesian approaches were used only to
solve the inverse problem of denoising the collected image. This
would correspond to estimating $I_p$ (i.e., it corresponds to the
photometric approaches proposed in \cite{ref:KVSL}) or to estimating $\zeta^*$
(the localization problem) from Eq.~\ref{eq:PoissonRate}.  Both of these approaches by
necessity already assume that a planet is present and typically take more time.   A hypothesis
testing approach is a direct approach that will more
appropriately be performed before any further processing.

\subsection{Odds ratio}
As the objective is not only to detect a potential planet but
also its discrete location $n^*$, we propose a test to be
performed on each pixel. To determine whether a potential planet
is present in the $n-th$ pixel, we compare the probabilities of
the following two alternate hypotheses:
\begin{eqnarray*}
H_{0}^n &:&\text{there is no planet in the }n-th\text{ pixel,} \\
H_{1}^n &:&\text{there is a planet in the }n-th\text{ pixel.}
\end{eqnarray*}
Associate a model $M_i$ to each hypothesis $H_i$:
\begin{eqnarray*}
M_{0}^n &:& \lambda_n=\lambda_{n}(I_b,I_p,\zeta^*),\quad \zeta^* \in [n\Delta\alpha, (n+1)\Delta\alpha],  \quad I_{p}=0\\
M_{1}^n &:& \lambda_n=\lambda_{n}(I_b,I_p,\zeta^*), \quad \zeta^*
\in [n\Delta\alpha, (n+1)\Delta\alpha], \quad I_{p}\neq 0
\end{eqnarray*}

 We now compare the posterior probabilities
of the two models conditioned on the collected data, and we shall
decide that there is a planet in pixel $n$ if the \emph{odds
ratio}, $O_{10}^n$, favors $H_{1}^n$ over $H_{0}^n$:
\begin{equation}
O_{10}^n =\frac{p\left( M_{1}^n|Z\right) }{p(M_{0}^n|Z)}=\frac{%
p\left( M_{1}^n\right) p\left( Z|M_{1}^n\right)
}{p(M_{0}^n)p\left( Z|M_{0}^n\right) }\label{defO10}
\end{equation}
where we have used Bayes' rule to write the odds ratio as the
product of the ratio of prior probabilities $\frac{p\left( M_{1}^n\right) }{%
p(M_{0}^n)}$ and the ratio of likelihoods $\frac{p\left( Z|M_{1}^n\right) }{%
p\left( Z|M_{0}^n\right) }$.  This dependence on the prior probability of a terrestrial planet in a given system (often referred to as $\eta_{\mbox{\scriptsize Earth}}$) is one of the most attractive features of the Bayesian approach, as information from prior missions such as Kepler can be explicitly included in the detection algorithm and the sensitivity of the detection to these assumptions can be quantitatively assessed.

One possible decision criterion would be that $O_{10}^n>1$
favors $H_1^n$, thus indicating that  there is a planet whose true location
$\zeta^*$ is inside the $n-th$ pixel.  Alternatively, the decision can be made based on a different value of the ratio derived from particular weighting criteria.  We call this a ``Loss-based decision'' and describe it next.

\subsection{Loss-based decision}
 As mentioned above, a decision criterion based on an odds ratio
of unity is not the only possible one.  In particular, an
alternative criteria uses a weighted decision based upon our
sensitivity to different types of errors (for instance, we may be
more willing to tolerate missed detections than false alarms).
To quantify this we define two possible actions:
\[
a_{0}:\text{ ``we accept }H_{0}"\quad a_{1}:\text{ ``we reject
}H_{0}".
\]
The Receiving Operator Curve (ROC) is used to determine the best possible decision threshold and quantify the relative importance of the two types of errors.  The ROC is a plot of the evolution of the missed
detection probability (Type II error), $p(a_0|H_{1}^{n})$, versus
$1-p(a_1|H_{0}^{n})$, where $p(a_1|H_{0}^{n})$ is the false alarm
rate (Type I error).  In other words, we place a point on the curve corresponding to the two probabilities for every possible threshold value, from 0 to infinity.  We then typically select the threshold corresponding to the ``knee'' of the curve, that is, the combination of maximum probability of success with minimum likelihood of missed detection.   The area under the ROC is defined as the ``power'' of the test and for the results below is parameterized by the integration time.  A perfect test is one with zero probability of missed detection and zero probability of false alarm, resulting in a power value (area under the ROC) of one.  The values of the probabilities for the curve are found through Monte Carlo simulations.\footnote{It was found empirically that the optimal threshold value strongly depends upon the shape of the PSF, with a unity  threshold best only for a square PSF.}

\subsection{Likelihoods and priors}

To compute $O_{10}^n$ as defined in Eq.~\ref{defO10}, we need to
assign the priors $p\left( M_{i}^n\right),\mbox{ }  i=1,2$ and
compute the likelihoods $p\left( Z|M_{i}^n\right), \mbox{ } i=1,2$.
In the absence of any other information, we assume here that the two models
are \emph{a priori} equiprobable at each pixel: $p\left( M_{0}^n\right) =\frac{1}{2}, \mbox{ } p\left( M_{1}^n\right) =\frac{%
1}{2}, \mbox{ } \forall n=1,\ldots,M$.

In general, the probability of a given data set depends on various physical parameters of the system. For instance, the mean photon rate in the Poisson distribution depends on the irradiance and the planet location in the pixel.  In order to compute the  likelihood, we must \emph{marginalize} the probability to remove this parameter dependence.  The marginalization law allows us to compute the global
likelihood of the model $M_{i}^n(\psi)$ over its parameter vector
 denoted by $\psi $:
\begin{equation}
p\left( Z|M_{i}^n\right) =\int d\psi p\left( \psi |M_{i}^n\right)
p\left( Z|\psi ,M_{i}^n\right).\label{eqmarg}
\end{equation}

Modeling each pixel intensity $z_j$ as an independent Poisson
random variable, we obtain
\begin{equation}
p\left( Z|\psi ,M_{i}^n\right)=\prod_{j=1}^{M}p\left( z_j|\psi
,M_{i}^n\right)=\prod_{j=1}^{M}\frac{{\lambda(\psi,M_{i}^n)_j}^{z_j}}{z_j!}\exp{-(\lambda(\psi,M_{i}^n)_j)}
\label{prodpoisson}
\end{equation}
which we multiply by the prior probabilities and then integrate over $\psi$ to determine the likelihoods as in Eq.~\ref{eqmarg}.

To determine the prior probabilities,  we assume that all the parameters are independent, resulting in:
\[
p\left( I_{p},I_{b},\zeta^* |M_{1}^n\right) =p\left(
I_{p}|M_{1}^n\right) p\left( I_{b}|M_{1}^n\right) p\left( \zeta^*
|M_{1}^n\right) ,
\]
where here $\psi$ consists of the planet and background irradiances and the planet location. 
We thus need to assign prior probabilities to each parameter individually. The
choice of specific densities for the priors is a delicate subject
as there is not necessarily a sound theoretical basis for the
choice and particular priors may effect the performance of the
estimator.  Our approach here is to make a sensible choice for
the priors that also makes the mathematical analysis manageable.
In future work we will examine the effect on the probability of success
when the assumed prior densities are actually in error.

The continuous planet location $\zeta^*$ is assumed to be uniform
over a pixel containing a planet:
\begin{eqnarray*}
 p\left( \zeta^* \right|M_1^n) &=&\frac{1}{\Delta
\alpha} \quad \text{when } \zeta^* \in [n\Delta\alpha,(n+1)\Delta\alpha], \\
&=&0 \quad \text{ otherwise.}
\end{eqnarray*}

As the intensity $I_{p}$ of the planet is strictly positive, it
has been modeled by a Gamma law $\Gamma \left( \alpha ,\beta
\right)$ of mean $E[I_p]=\alpha\beta$ and variance
$E[I_p^2]=\alpha \beta^2$.  A second advantage of this model is that the gamma distribution is the
conjugate prior of a Poisson law (\cite{Berger}), thus simplifying
future computations:
\begin{equation}
p\left( I_{p}\right|M_1^n) =I_{p}^{\alpha -1}\exp \left( -\frac{I_{p}}{\beta }%
\right) .\label{eqgamma}
\end{equation}

We will either assume that the intensity of the background $I_{b}$
is known or that it also follows a Gamma law, $\Gamma \left(
a,b\right)$.

\section{Results}

For simplicity, all the results provided in this section have been performed for a 1-D point spread function, that is, a cross section of the two-dimensional image.  This function for optimal shaped pupil coronagraphs is the transform of the prolate spheroidal wavefunction (see \cite{ref:KVSL} or \cite{VSK02}). In future work we will repeat and compare to a number
of different pupil designs (as well as other types of
coronagraphs) and extend to two-dimensions.  All results are based on Monte-Carlo simulations.

To account for the size of the PSF, which extends over more than
one pixel, we define $\Delta $ as the width of the significant
part of the PSF (normally taken equal to the FWHM), and set
\[
K=ceil\left( \frac{\Delta }{\Delta \alpha }\right)
\]
such that $2K+1$ is the maximum number of pixels that can be
covered by the truncated PSF and $\Delta S =(2K+1)\Delta \alpha$.
For a pixel $n_j$ such that $\|n_j-n\|>K$,
$\lambda(\psi,M_{0}^n)_j=\lambda(\psi,M_{1}^n)_j=I_b \Delta \alpha
\tau$ . The ratio of likelihoods as defined in
Eq.~\ref{prodpoisson} only involves the pixels covered by $\Delta$:
\begin{eqnarray*}
\frac{p\left(Z|\psi,M_{1}^n\right)}{p\left(Z|\psi,M_{0}^n\right)}&=&\frac{\prod_{j=1}^{M}p\left(
z_j|\psi ,M_{1}^n\right)}{\prod_{j=1}^{M}p\left( z_j|\psi
,M_{0}^n\right)}\\
&=&\prod_{j=\max(n-K,1)}^{\min(n+K,M)}\frac{p\left(z_j|\psi
,M_{1}^n\right)}{p\left( z_j|\psi
,M_{0}^n\right)}\\
&=&\prod_{j=\max(n-K,1)}^{\min(n+K,M)}
\left(\frac{{\lambda(\psi,M_{1}^n)_j}}{{\lambda(\psi,M_{0}^n)_j}}\right)^{z_j}\exp{-(\lambda(\psi,M_{1}^n)_j-\lambda(\psi,M_{0}^n)_j)}
\end{eqnarray*}
where $M$ is the total number of pixels.

In the following simulations we have selected the pixel size such
that $K=2$, and for the sake of simplicity, we will consider $n$
such that $\max(n-K,1)= n-K$, $\min(n+K,M)=n+K$ (no edge).

\subsection{Assume $I_{p}$ and $I_{b}$ are known}
For this study, we use Monte-Carlo simulations to determine
the effectiveness of the hypothesis testing criterion and the resulting probabilities of type I and II errors for a given normalized integration time, $\tau$.   This can then be compared to the decision criterion described in \cite{ref:KVSL} and based on Eq.~\ref{eqsn} for $t_1=\tau$.  In this section, the simulations assume $I_p$ and $I_b$ are both known.   In
reality, of course, not only is $I_p$ not necessarily known, it
varies with pixel location as the planet moves further away from
the star (assuming a reflected visible light system).  This assumption is relaxed in later sections.

Under these assumptions, the odds ratio in favor of $H_1^n$ is
(see Appendix):
\begin{equation}
O_{10}^n =\frac{1}{\Delta \alpha }\exp \left( -\tau I_{p}\Delta
P\right) \int_{0}^{\Delta \alpha }\prod_{j=n-K}^{j=n+K}\left(
1+\frac{Q}{\Delta \alpha}\Delta P_j\left( \theta \right) \right)
^{z_{j}}d\theta ,
\end{equation}\label{eq1}
where $Q=\frac{I_p}{I_b}$ and $\Delta P_j(\theta)$ is the area of
the PSF covered by one of the five pixels (indexed by $j$) when the
planet appears at location $\theta$ in pixel $n$. Figure 1
depicts the results obtained for two simulated sets of images.
For $i\in[1,500]$, the data are simulated without any planet
($Q=0$), while for $i\in[501,1000]$, $Q=1$, the planet is located
in pixel $n^*=11$.  For all simulations, the normalized
integration time was set to $\tau=1$, corresponding to a
signal-to-noise ratio $S/N$ computed as in Eq.~\ref{eqsn} of
$3.52$.  While it is reasonable to expect some detections using the conventional approach with this $S/N$ ratio, using the usual criteria of $S/N=5$ described above,  no planets
would be detected.  Figure~2 depicts the histogram of $\log
O_{10}^{n^*}$ for the two simulated sets. For a threshold equal
to one, the false alarm rate is $0.1\%$, while the missed detection
rate reaches $7.2\%$. From the ROC curve (not plotted here for
brevity), we determined that the test has maximal power if the
threshold is set to $0.158$ rather than 1.  With this threshold,
the false alarm rate is $4.0\%$ and the missed detection rate is
$3.0\%$.   This demonstrates that under the same assumptions, the Bayesian technique is extremely effective at detection with quantifiable performance and at shorter integration times.

\begin{figure}[h]
\parbox{6cm}{\includegraphics[width=2.5in,height=2in]{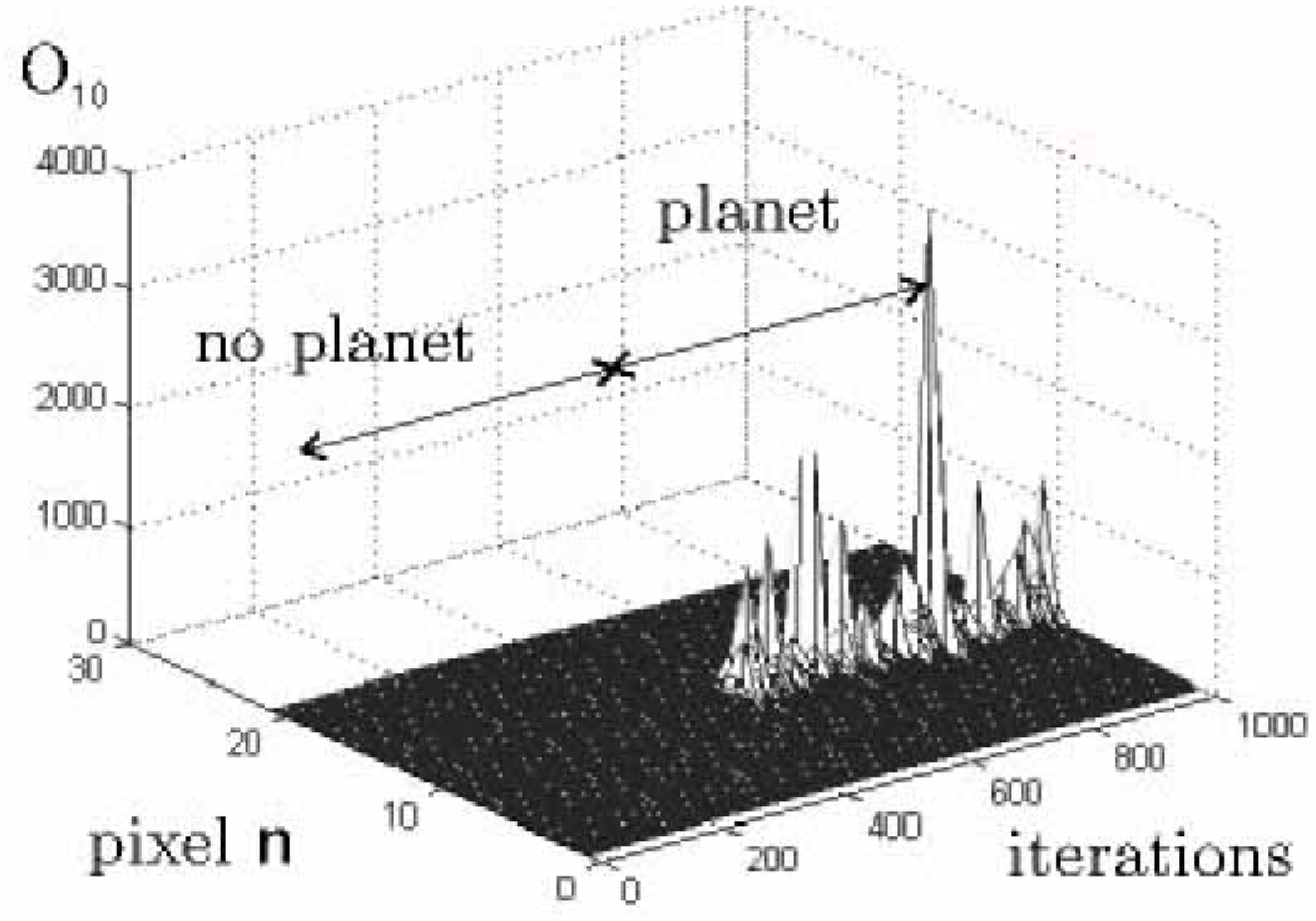}
\caption{Value of $O_{10}^n$ as a function of $n$ for the two
simulated sets, for $\tau =1$.}
}\parbox{3cm}{\hspace{3cm}}\parbox{5cm}{\includegraphics[width=2in,height=2in]
{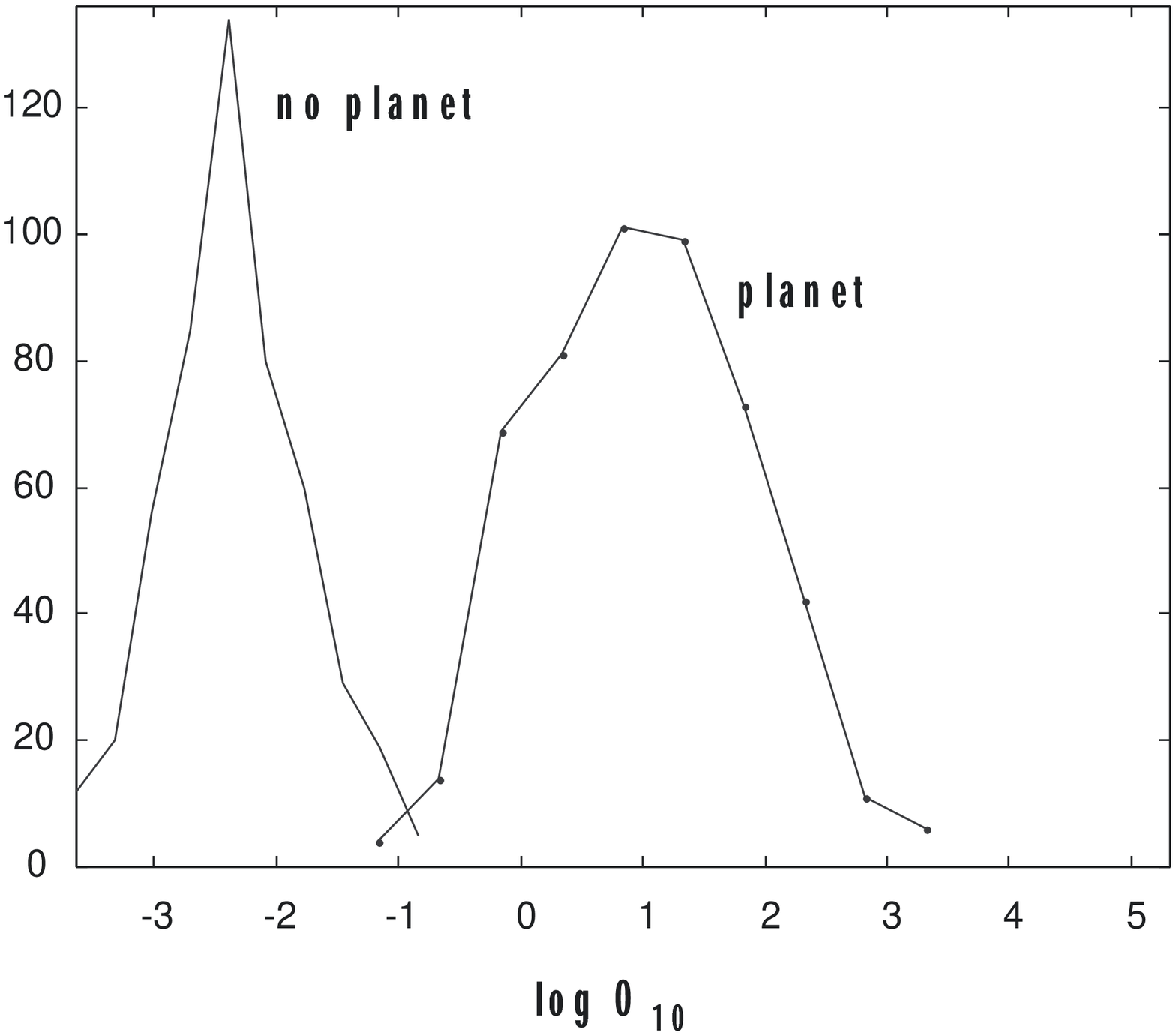}
 \caption{Histogram of $\log O_{10}^{n^*}$ for the two simulated sets, $\tau=1$}}
\end{figure}

\subsection{Assume $I_{p}$ is uncertain}

The first complication to consider is a lack of knowledge of $I_p$. 
 In \cite{ref:KVSL}, we take
the approach of estimating $I_p$ (whether a planet is there or not) and examining the quality of the
estimate to make the decision.  Here, we avoid photometry and
still use a Bayesian hypothesis testing approach to make the
decision in a shorter integration time.

If we assume under $H_1$ that $I_p$ follows a gamma law,
$\Gamma(\alpha,\beta)$, then using Eqs.~\ref{defO10} and \ref{eqmarg}
we get
\begin{equation}
O_{10}^n =\frac{1}{\Delta \alpha }
\sum_{p=0}^{p=\sum_{j=n-K}^{n+K} z_j }\frac{\Gamma \left(
\alpha +p\right) }{\Gamma \left( \alpha \right) }\left( \frac{\beta }{I_{b}}%
\right) ^{p}\frac{1}{(1+\tau \beta \Delta
P)^{\alpha+p}}\int_{0}^{\Delta \alpha }\gamma _{p}\left( \theta
\right) d\theta
\end{equation}
where the coefficients $\gamma _{p}(\theta)$ are provided in the
Appendix. The data have been simulated for $\tau=1$, with
$\alpha=4$ and $\beta=QI_b/4$. The histogram of the
resulting estimated $S/N$ using Eq.~\ref{eqsn} and the $I_p$
selected from the distribution from each run is depicted on
Figure~3.  Following Eq.~\ref{eqsn}, if a conclusion is reached
only for $S/N>5$, the missed detection rate reaches 72.7\%.  As described in Section 3.1, the mean value of $I_p$ was chosen to correspond to a $S/N$ of 3.52 for $\tau=1$.  When $I_p$ is sampled from the gamma distribution here (with the same mean value), roughly 24\% of the time a planet occurs with a mean $S/N>5$ that would assure a detection by the conventional approach.
For the same simulated data, our
criterion provides a missed detection rate of only 1.0\% with the odds
ratio decision threshold set at one.

\begin{figure}[h]
\parbox{6cm}{\includegraphics[width=2.7in,height=1.8in]{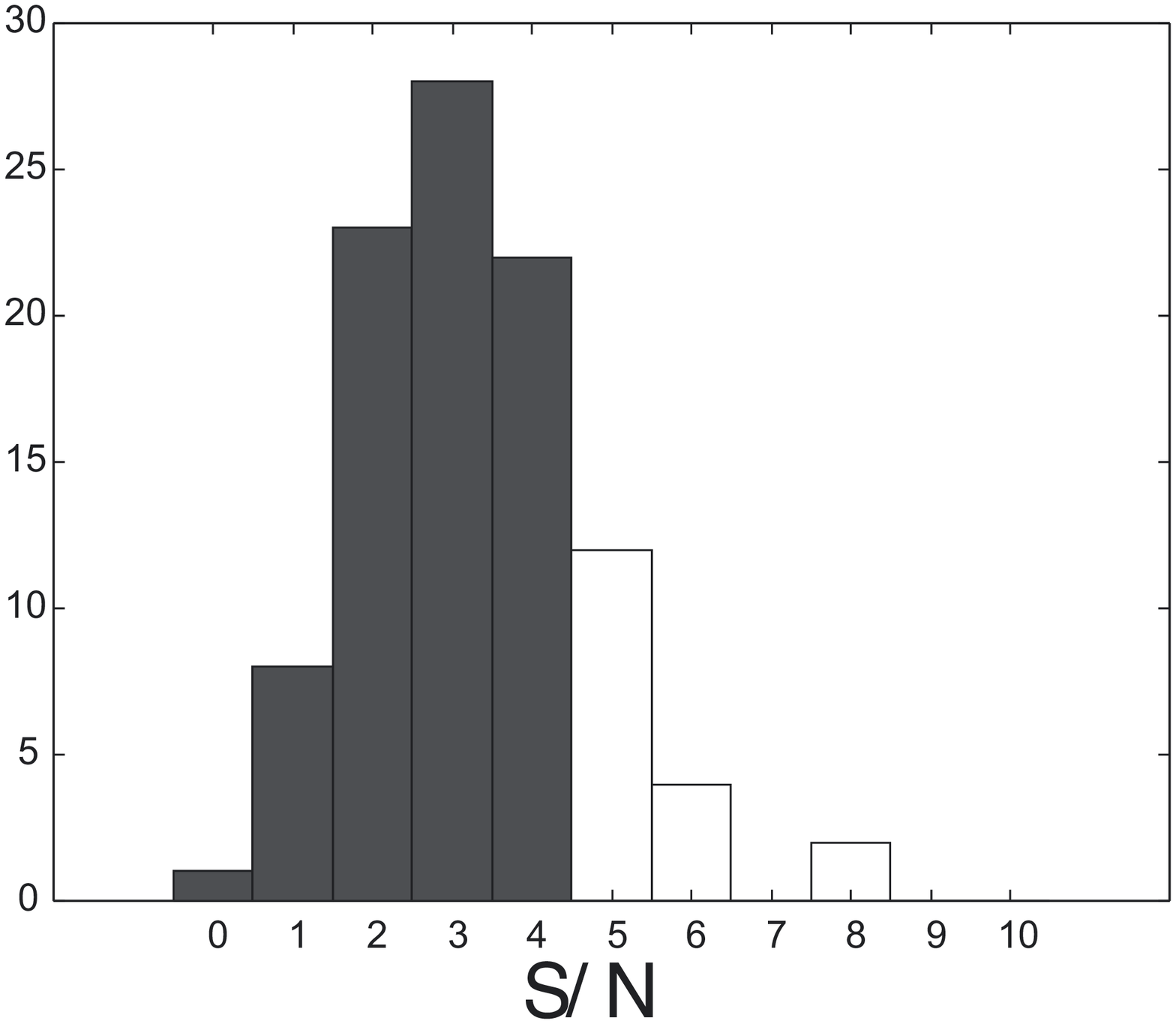}
\caption{Histograms of the estimated $S/N$ when the planet intensity
$I_p$ follows a Gamma law. The blue boxes correspond to the case
where $S/N < 5$, \emph{i.e.} that an existing planet has not been
detected.}}
\parbox{3cm}{\hspace{2cm}}\parbox{5cm}{\includegraphics[width=2.2in,height=1.9in]{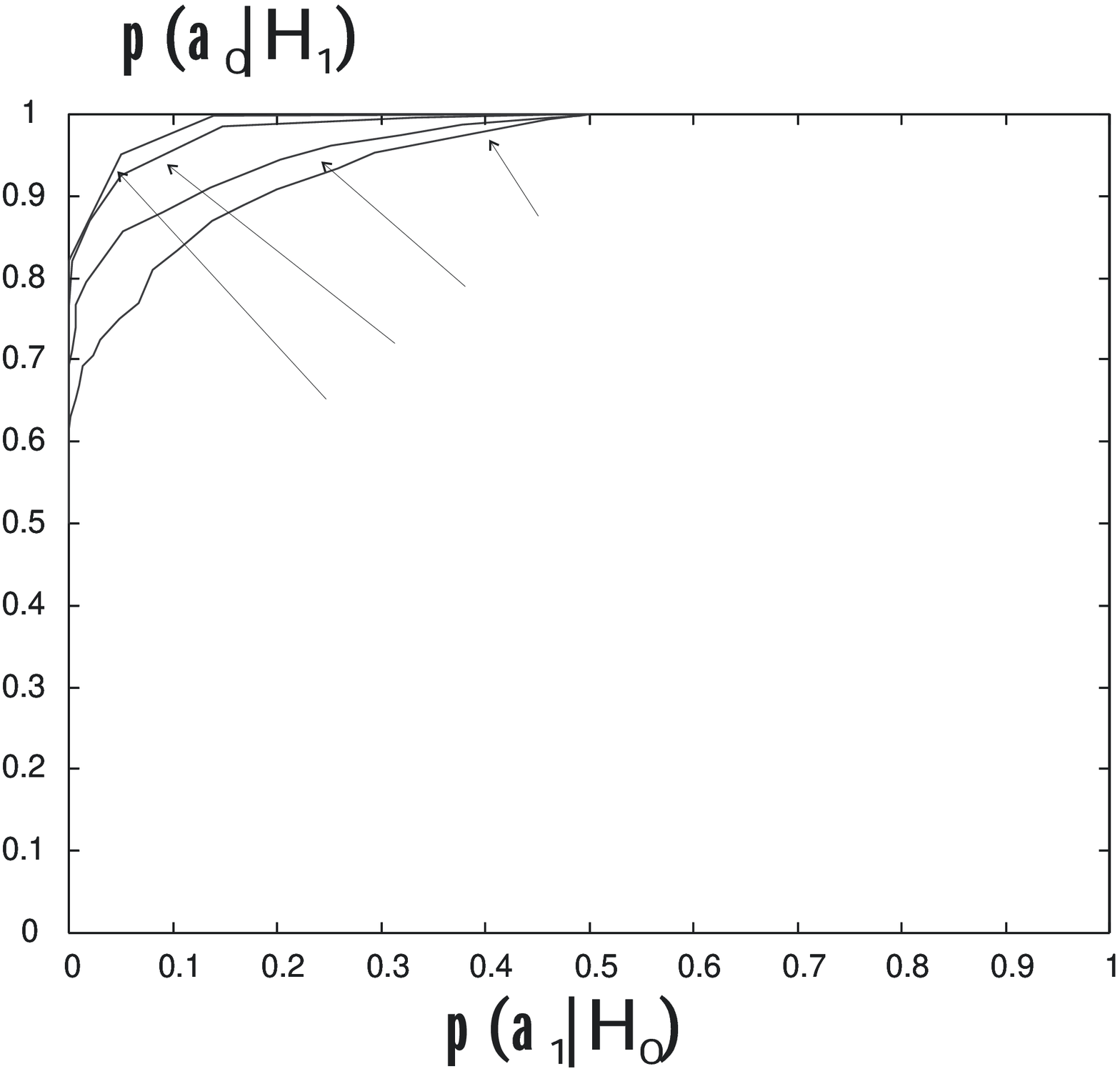}
\caption{ROC curves as a function of the relative integration
time $\tau$: the power of the test increases with $\tau$}}
\end{figure}

The Receiver Operating Curves  for this case is depicted in
Figure~4.  It shows the trade-off between the false alarm and the
missed detection rates for different values of the relative
integration time $\tau$.  As expected, the greater $\tau$, the
more powerful the test. Nevertheless, note that a relatively
high correct decision probability $p(a_0|H_0)+p(a_1|H_1)$ (found
via the Monte Carlo results with a decision threshold of one) can
be achieved even for low integration times (see Table~1). In practice on real data we may choose the lower optimal
threshold deduced from the simulations, thus achieving slightly
higher probabilities of success.

\begin{table}[h]
\begin{center}
\begin{tabular}{|c|c|c|c|c|} \hline
$\tau$ & $0.1$ & $0.2$ & $0.5$ & $1$\\
\hline
$p(a_0|H_0)+p(a_1|H_1)$ & $0.732$ & $0.804$ & $0.876$ & $0.900$\\
\hline
\end{tabular}
\caption{Evolution of the correct decision probability with
$\tau$}
\end{center}
\end{table}

\subsection{Assume $I_{b}$ is also unknown}
The final, and most complicated, case is to assume that both $I_p$
and $I_b$ are unknown.  Again, in our previous report and in
\cite{ref:KVSL}, we resorted to two-parameter least-squares
estimation (matched filtering) on the photometric data.  By
simultaneously estimating the irradiance of the planet and the
background, a decision can be made on the planet, but the
integration times become quite long.  With the Bayesian approach, we found that we have a
very high probability of success even at the same integration
time as the previous case.

For the sake of brevity, we defer the derivation to the Appendix. If $p(I_{b})=\Gamma(a,b)$, we use the
marginalization law Eq.~\ref{eqmarg} to eliminate its dependence
from the odds ratio and obtain:
\[
O_{10}^n =\frac{1}{\Delta \alpha }\frac{1}{\left( 1+\beta \tau
\Delta P\right)^{\alpha+p}}\sum_{p=0}^{p=\sum_{j=n-K}^{n+K} z_j
}\frac{\Gamma
\left( \alpha +p\right) }{\Gamma \left( \alpha \right) }\left(\frac{\beta^{\prime}}{%
b^{\prime}}\right)^p\frac{\Gamma \left( \sigma_K-p\right) }{%
\Gamma \left(  \sigma_K+a\right) }\int_{0}^{\Delta \alpha }\gamma
_{p}\left( \theta \right) d\theta
\]
where
\[
\frac{1}{b ^{\prime}} =\frac{1}{b}+\tau \Delta S
\] 
A similar Monte-Carlo study as before shows that we can reach a correct decision probability of 0.84 for the same integration time (
$\tau=1$) with $\frac{E[I_p]}{E[I_b]}=1$ and $E[I_p^2]=E[I_b]/2,
E[I_b^2]=E[I_b]/2$.

\section{Conclusion}
Bayesian hypothesis testing is  a promising approach to planet detection and to
associate a quantitative confidence level with the detection
system, with a relative integration time smaller than $t_1$ and
without performing photometry.  Moreover this approach allows us to
take into account the uncertainty associated with the nuisance
parameters such as the intensity of the background via the
marginalization law. Nevertheless, choosing a prior may strongly
influence the quality of the result. Further work aims to develop
this same approach with other priors and to study its robustness.

No speckle has been considered yet. Current work aims at defining
a three-hypothesis test to detect if the pixel under study
contains a planet, speckle, or just noise. A trio of three
bandpass filters, while allowing us to consider
quasi-monochromatic signals (and directly applying this
technique), would also permit us to define speckle as a
space-variant component.  Note that adding other features in addition to just
the PSF shape in a Bayesian or optimal classifier could also
improve the results.


\newpage

\appendix

\section{Appendix}

\subsection{PSF Truncation}

Let $\Delta $ be the FWHM of the PSF. The maximum number of pixels that can
be covered by the PSF is an even number denoted by $2K+1$ where
\[
K=ceil\left( \frac{\Delta }{\Delta \alpha }\right) .
\]
As
\[
p\left( z_{k}|M_{1}^{n}\right) =p\left( z_{k}|M_{0}\right) ,k\notin \left[
n-K,n+K\right]
\]
Then

\[
O_{10}^{n}=\frac{\prod\limits_{k=1}^{M}p\left( z_{k}|M_{1}^{n}\right) }{%
\prod\limits_{k=1}^{M}p\left( z_{k}|M_{0}^{n}\right) }=\frac{%
\prod\limits_{k=n-K}^{n+K}p\left( z_{k}|M_{1}^{n}\right) }{%
\prod\limits_{k=n-K}^{n+K}p\left( z_{k}|M_{0}^{n}\right) }\triangleq \frac{%
L_{1}\left( I_{p},I_{b},\theta \right) }{L_{0}\left( I_{b}\right) }
\]

\textbf{Example.}\emph{ If $K=1,$ 3 pixels only can be involved,}
\[
O_{10}^{n}=\frac{p\left( z_{n-1}|M_{1}^{n}\right) p\left(
z_{n}|M_{1}^{n}\right) p\left( z_{n+1}|M_{1}^{n}\right) }{p\left(
z_{n-1}|M_{0}^{n}\right) p\left( z_{n}|M_{0}^{n}\right) p\left(
z_{n+1}|M_{0}^{n}\right) }
\]

In the following, we will use the following notation
\[
\sigma _{K}\triangleq \sum\limits_{k=n-K}^{k=n+K}z_{k};\pi _{k}
\triangleq \prod\limits_{k=n-K}^{n+K}z_{k}!
\]

\subsection{Ip is assumed to be known}

Obviously, as $I_{b}$ is assumed to be known
\[
p\left( Z|M_{0}^{n}\right) = \int_{\Bbb{R}}p\left(
Z|M_{0}^{n},I_{b}\right) p\left( I_{b}\right) dI_{b}=p\left(
Z|M_{0}^{n},I_{b}\right) .
\]
For the constant-rate model $M_{0}^{n}$

\begin{eqnarray*}
&&L_{0}\left( I_{b}\right) =\prod\limits_{k=n-K}^{n+K}p\left(
z_{k}|M_{0}^{n},I_{b}\right) =\prod\limits_{k=n-K}^{n+K}\frac{1}{z_{k}!}%
\lambda _{k}^{z_{k}}\exp \left( -\lambda _{k}\right)  \\
&=&\prod\limits_{k=n-K}^{n+K}\frac{1}{z_{k}!}\left( I_{b}\Delta \alpha \tau
\right) ^{z_{k}}\exp \left( -I_{b}\Delta \alpha \tau \right) =\frac{1}{\pi
_{K}}\left( I_{b}\tau \Delta \alpha \right) ^{\sigma _{K}}\exp \left(
-I_{b}\tau \Delta S\right) ,
\end{eqnarray*}
where $\Delta S=\left( 2K+1\right) \Delta \alpha$ is the area of
integration around the $n^{\prime }th$ pixel.

Assuming that $\theta $ is uniform on $\left[ \zeta _{n},\zeta _{n+1}\right]
$, the likelihood of the alternative model $M_{1}^{n}$ is
\begin{eqnarray*}
L_{1}\left( I_{p},I_{b}\right)  &=&\int_{\zeta _{n}}^{\zeta _{n+1}}p\left( Z|M_{1}^{n},I_{b},I_{p},\theta
\right) p\left( \theta \right) d\theta  \\
&=&\frac{1}{\Delta \alpha }\int_{\zeta _{n}}^{\zeta _{n+1}}d\theta
\prod\limits_{k=n-K}^{n+K}p\left( z_{k}|M_{1}^{n},I_{b},I_{p},\theta \right)
\\
&=&\frac{1}{\Delta \alpha }\int_{\zeta _{n}}^{\zeta _{n+1}}d\theta \frac{1}{%
\pi _{K}}\prod\limits_{k=n-K}^{n+K}\left( I_{b}\Delta \alpha \tau
+I_{p}\Delta P_{k}\left( \theta \right) \right) ^{z_{k}}\exp \left( -\tau
\left( I_{b}\Delta \alpha +I_{p}\Delta P_{k}\left( \theta \right) \right)
\right)
\end{eqnarray*}

Compute
\begin{eqnarray*}
A_{1}\left( \theta \right)  &=&\prod\limits_{k=n-K}^{n+K}\left( I_{b}\Delta
\alpha \tau +\tau I_{p}\Delta P_{k}\left( \theta \right) \right) ^{z_{k}}\exp
\left( -\tau \left( I_{b}\Delta \alpha +I_{p}\Delta P_{k}\left( \theta
\right) \right) \right)  \\
&=&\left( I_{b}\tau \Delta \alpha \right) ^{\sigma _{K}}\exp \left(
-I_{b}\tau \Delta S\right) \prod\limits_{k=n-K}^{n+K}\left( 1+\frac{%
I_{p}\Delta P_{k}\left( \theta \right) }{I_{b}\tau \Delta \alpha }\right)
^{z_{k}}\exp \left( -\tau \sum\limits_{k=n-K}^{k=n+K}I_{p}\Delta P_{k}\left(
\theta \right) \right)  \\
&=&\left( I_{b}\tau \Delta \alpha \right) ^{\sigma _{K}}\exp \left( -\tau
\left( I_{b}\Delta S+I_{p}\Delta P\right) \right) \prod\limits_{k=n-K}^{n+K}\left( 1+Q\frac{\Delta P_{k}\left(
\theta \right) }{\Delta \alpha }\right) ^{z_{k}}
\end{eqnarray*}
by noting that $\sum\limits_{k=n-K}^{k=n+K}\Delta P_{k}\left( \theta \right)
=\Delta P$ and by noting $Q=\frac{I_{p}}{I_{b}}$. We then have
\[
L_{1}\left( I_{p},I_{b}\right) =\frac{\left( I_{b}\tau \Delta \alpha \right)
^{\sigma _{K}}}{\Delta \alpha \pi _{K}}\exp \left( -\tau \left( I_{b}\Delta
S+I_{p}\Delta P\right) \right) \int_{\zeta _{n}}^{\zeta _{n+1}}d\theta
\prod\limits_{k=n-K}^{n+K}\left( 1+Q\frac{\Delta P_{k}\left( \theta \right)
}{\Delta \alpha }\right) ^{z_{k}}
\]
\begin{eqnarray*}
O_{10}^{n}\left( I_{p},I_{b}\right)  &=&\frac{L_{1}\left( I_{p},I_{b}\right)
}{L_{0}\left( I_{b}\right) } \\
&=&\frac{1}{\Delta \alpha }\exp \left( -\tau I_{p}\Delta P\right)
\int_{\zeta _{n}}^{\zeta _{n+1}}d\theta \prod\limits_{k=n-K}^{n+K}\left( 1+Q%
\frac{\Delta P_{k}\left( \theta \right) }{\Delta \alpha }\right) ^{z_{k}}
\end{eqnarray*}

\subsection{Assume $I_{p}$ is uncertain}

As $I_{b}$ is still assumed to be known, $L_{0}$ is unchanged. Using the
marginalization law, we have
\begin{eqnarray*}
O_{10}^{n}\left( I_{b}\right)  &=&\frac{\int_{0}^{\infty} p\left(
I_{p}\right)
L_{1}\left( I_{p}\right) dI_{p}}{L_{0}\left( I_{b}\right) } \\
&=&\frac{1}{\Delta \alpha }\int dI_{p}p\left( I_{p}\right) \exp \left( -\tau
I_{p}\Delta P\right) \int_{\zeta _{n}}^{\zeta _{n+1}}d\theta
\prod\limits_{k=n-K}^{n+K}\left( 1+\frac{I_{p}}{I_{b}}\frac{\Delta
P_{k}\left( \theta \right) }{\Delta \alpha }\right) ^{z_{k}} \\
&=&\frac{1}{\Delta \alpha \Gamma \left( \alpha \right) \beta ^{\alpha }}%
\int_{\zeta _{n}}^{\zeta _{n+1}}d\theta \int_{0}^{\infty}
dI_{p}I_{p}^{\alpha -1}\exp \left( -\frac{I_{p}}{\beta ^{\prime
}}\right) \prod\limits_{k=n-K}^{n+K}\left( 1+I_{p}\frac{\Delta
P_{k}\left( \theta \right) }{I_{b}\Delta \alpha }\right) ^{z_{k}},
\end{eqnarray*}
by posing
\[
\frac{1}{\beta ^{\prime }}=\frac{1}{\beta }+\tau \Delta P
\]
For the sake of simplicity, we only present here the case for $K=2$ (case
described in the main document).

Using the binomial law
\[
\left( 1+\alpha x\right) ^{\beta }=\sum_{n=0}^{\beta }\binom{\beta }{n}
\alpha ^{n}x^{n}
\]
$2K+1$ times we have that
\[
\prod\limits_{k=n-K}^{n+K}\left( 1+I_{p}\frac{\Delta P_{k}\left( \theta
\right) }{I_{b}\Delta \alpha }\right) ^{z_{k}}=\sum_{p=0}^{\sigma
_{K}}\gamma _{p}\left( \theta \right) \left (\frac{I_{p}}{I_{b}}\right)^{p}
\]
where
\begin{eqnarray*}
\gamma _{p}\left( \theta \right)  &=&\sum_{r=\max \left( 0,p-z_{n+1}\right)
}^{\min \left( \sum_{n-K}^{n}z_{j},p\right) }\beta _{r}\left( \theta \right)
\binom{z_{n+1}}{p-r}\left( \frac{\Delta P_{n+1}\left( \theta \right) }{
\Delta \alpha }\right) ^{p-r}; \\
\beta _{r}\left( \theta \right)  &=&\sum_{l=\max \left(
0,r-z_{n}-z_{n+1}\right) }^{\min \left( z_{n-K}+z_{n-K+1},r\right) }\alpha
_{l}\left( \theta ,n,n+1\right) \alpha _{p-l}\left( \theta ,n+2,n+3\right)
\\
\alpha _{l}\left( \theta ,i,j\right)  &=&\sum_{m=\max \left(
0,l-z_{j}\right) }^{\min \left( z_{i},l\right) }\binom{z_{i}}{m}\binom{z_{j}
}{l-m}\left( \frac{\Delta P_{i}\left( \theta \right) }{\Delta \alpha }
\right) ^{m}\left( \frac{\Delta P_{j}\left( \theta \right) }{\Delta \alpha }
\right) ^{l-m}
\end{eqnarray*}
so that
\begin{eqnarray*}
O_{10}^{n}\left( I_{b}\right)  &=&\frac{1}{\Delta \alpha \Gamma \left(
\alpha \right) \beta ^{\alpha }}\int_{\zeta _{n}}^{\zeta _{n+1}}d\theta
\int_{0}^{\infty }dI_{p}I_{p}^{\alpha -1}\exp \left( -\frac{I_{p}}{\beta
^{\prime }}\right) \sum_{p}\gamma _{p}\left( \theta \right) \left( \frac{%
I_{p}}{I_{b}}\right) ^{p} \\
&=&\frac{1}{\Delta \alpha \Gamma \left( \alpha \right) \beta ^{\alpha }}%
\sum_{p}\frac{1}{I_{b}^{p}}\int_{\zeta _{n}}^{\zeta _{n+1}}\gamma _{p}\left(
\theta \right) d\theta \int_{0}^{\infty }dI_{p}I_{p}^{p+\alpha -1}\exp
\left( -\frac{I_{p}}{\beta ^{\prime }}\right)  \\
&=&\frac{1}{\Delta \alpha \Gamma \left( \alpha \right) \beta ^{\alpha }}%
\sum_{p}\frac{1}{I_{b}^{p}}\int_{\zeta _{n}}^{\zeta _{n+1}}\gamma _{p}\left(
\theta \right) d\theta \Gamma \left( \alpha +p\right) \beta ^{\prime
^{\alpha +p}}
\end{eqnarray*}

so
\[
O_{10}^{n}\left( I_{b}\right) =\frac{1}{\Delta \alpha }\sum_{p}\frac{\Gamma
\left( \alpha +p\right) }{\Gamma \left( \alpha \right) }\left( \frac{1}{%
1+\tau \Delta P \beta}\right) ^{\alpha +p}\left( \frac{\beta }{I_{b}}\right)
^{p}\int_{0}^{\Delta \alpha }\gamma _{p}\left( \theta \right) d\theta .
\]

\textbf{Remark.}\emph{ Note that the computation of
$\int_{0}^{\Delta \alpha }\gamma _{p}\left( \theta \right)
d\theta $, that may look very time-consuming, can be done
before any data are collected. To be computed, only the sums (depending on $%
z_{i}$) have to be computed online.}

\subsection{When I$_{b}$ is uncertain}

We now have to apply the marginalization law for both sides.

\begin{eqnarray*}
L_{0} &=&\int_{0}^{\infty }p\left( I_{b}\right) L_{0}\left( I_{b}\right)
dI_{b} \\
&=&\frac{\left( \tau \Delta \alpha \right) ^{\sigma _{K}}}{\Gamma \left(
a\right) b^{a}\pi _{K}}\int_{0}^{\infty }I_{b}^{\sigma _{K}+a-1}\exp \left(
-\frac{I_{b}}{b'} \right) dI_{b} \\
&=&\frac{\left( \tau \Delta \alpha \right) ^{\sigma _{K}}}{\pi _{K}}\frac{%
\Gamma \left( a^{\prime }\right) b^{\prime a^{\prime }}}{\Gamma \left(
a\right) b^{a}} \\
&=&\frac{\left( \tau \Delta \alpha \right) ^{\sigma _{K}}}{\pi _{K}}\frac{%
\Gamma \left( a+\sigma _{K}\right) }{\Gamma \left( a\right) }\frac{b^{\prime
a^{\prime }}}{b^{a}}
\end{eqnarray*}

where
\[
a^{\prime }=\sigma _{K}+a;\frac{1}{b^{\prime }}=\frac{1}{b}+\tau \Delta S
\]

Compute now $L_{1}$%
\begin{eqnarray*}
L_{1} &=&\int_{0}^{\infty }p\left( I_{b}\right) L_{1}\left( I_{b}\right)
dI_{b} \\
&=&\frac{1}{\pi _{K}}\frac{1}{\Gamma \left( a\right) b^{a}}\int_{0}^{\infty
}I_{b}^{a-1}\exp \left( -\frac{I_{b}}{b}\right) \frac{1}{\Delta \alpha
\Gamma \left( \alpha \right) \beta ^{\alpha }}\sum_{p}\frac{\Gamma \left(
\alpha +p\right) \beta ^{\prime ^{\alpha +p}}}{I_{b}^{p}} \\
&&\int_{\zeta _{n}}^{\zeta _{n+1}}\gamma _{p}\left( \theta \right) d\theta
\left( I_{b}\tau \Delta \alpha \right) ^{\sigma _{K}}\exp \left( -I_{b}\tau
\Delta S\right) dI_{b} \\
&=&\frac{1}{\pi _{K}}\frac{1}{\Gamma \left( a\right) b^{a}}\frac{\left( \tau
\Delta \alpha \right) ^{\sigma _{K}}}{\Delta \alpha \Gamma \left( \alpha
\right) \beta ^{\alpha }} \\
&&\sum_{p}\Gamma \left( \alpha +p\right) \beta ^{\prime ^{\alpha
+p}}\int_{\zeta _{n}}^{\zeta _{n+1}}\gamma _{p}\left( \theta \right) d\theta
\int_{0}^{\infty }\exp \left( -\frac{I_{b}}{b^{\prime }}\right)
I_{b}^{a^{\prime \prime}-1}dI_{b} \\
&=&\frac{1}{\pi _{K}}\frac{1}{\Gamma \left( a\right) b^{a}}\frac{\left( \tau
\Delta \alpha \right) ^{\sigma _{K}}}{\Delta \alpha \Gamma \left( \alpha
\right) \beta ^{\alpha }}\sum_{p}\int_{\zeta _{n}}^{\zeta _{n+1}}\gamma
_{p}\left( \theta \right) d\theta \Gamma \left( \alpha +p\right) \beta
^{\prime ^{\alpha +p}}b^{\prime a^{\prime \prime }}\Gamma \left( a^{\prime
\prime }\right)
\end{eqnarray*}

where
\[
a^{\prime \prime }=\sigma _{K}+a-p=a^{\prime }-p;
\]

so
\[
O_{10}^{n} =\frac{1}{\Delta \alpha }\left( \frac{1}{1+\tau \beta
\Delta P}\right)
^{\alpha }\sum_{p}\left( \frac{\beta ^{\prime }}{b^{\prime }}\right) ^{p}%
\frac{\Gamma \left( \alpha +p\right) }{\Gamma \left( \alpha \right) }\frac{%
\Gamma \left( \sigma_K-p\right) }{\Gamma \left( \sigma_K+a\right) }%
\int_{\zeta _{n}}^{\zeta _{n+1}}\gamma _{p}\left( \theta \right) d\theta
\]

\end{document}